# Micro-Bioreactor Mimicking a Cellular Environment

**M. Hase, A. Yamada, T. Hamada, and K. Yoshikawa [†,a)]**

*Department of Physics, Graduate School of Science, Kyoto University, Kyoto 606-8502, Japan, and*

[†] *Spatio-Temporal Order Project, ICORP,JST*

a) Electronic mail: yoshikaw@scphys.kyoto-u.ac.jp

**Abstract**

We report that a cell-sized water droplet (CWD) with a diameter of several tens of microns can serve as a micro-biochemical reactor. Since the droplet inner surface is covered by a phospholipid layer, it provides an environment similar to that in living cells. The CWD is manipulated with laser tweezers and we monitor the time-development of biochemical reactions in a single CWD by fusing two droplets which contain a substrate and an enzyme, respectively. As actual examples of enzymatic reactions, we present results on the reaction of calcein with esterase and on the expression reaction of green fluorescent protein.

Living organisms on Earth maintain their vital activity by using micro-closed spatial units, i.e., a cellular structure. More than half a century ago, Oparin systematically studied the self-organization of a closed cell-like structure,[1] i.e., a coacervate, and found that several biochemical reactions can proceed within a coacervate. Several decades after Oparin published his monograph, a hypothesis was established that the main component of cellular membranes is phospholipids and various functional proteins are embedded in the phospholipid bilayer membrane.[2,3] Stimulated by this hypothesis, model membranes with phospholipids have been extensively and actively studied up to the present,[4-6] as a simple model of living cells. However, most previous investigations on closed phospholipid membranes have been performed for vesicles smaller than a scale of 100 nm, which is more than one order of magnitude smaller than the scale of living cells. Although several studies on cell-sized phospholipid vesicles larger than a scale of μm have been reported,[7-9] they used experimental methodologies with serious inherent problems, such as nonphysiological ionic conditions, difficult to encapsulate biological macromolecules without denaturation, difficult to fuse desired vesicles, etc.[10-13]

Recently, studies on the cell-sized water droplets (CWDs) in an oil phase stabilized by the presence of a surfactant such as span80 or tween20 have been carried out in a closed aqueous compartment on a micrometer scale, with an eye toward their use as cell models, as micro reactors, and in biochemical or medical applications.[14-17] CWDs are spontaneously formed through simple mixing of oil/water/surfactant, and these CWDs maintain their spherical geometries under mechanical and osmotic stresses. These advantages of CWDs allow experiments to be conducted under physiological ionic conditions and biological macromolecules can be easily encapsulated. The observation of biochemical reactions in CWDs has been reported, e.g. GFP expression,[14] PCR[15] etc. However, it is still rather difficult to monitor a desired biochemical reaction in a macro-reactor or CWD under real-time observation. Recently, we found that CWDs surrounded by phospholipid layers can be prepared by a simple mixing procedure.[18,19] Such CWDs may be suitable models of living cells, since phospholipid molecules are arranged on the surface with their hydrophilic moieties oriented toward the inner aqueous phase, as in the cytoplasmic membrane.

1,2-Dioleoyl-sn-glycero-3-phosphoethanolamine (DOPE), calcein acetoxymethyl ester (calcein-AM), and Fura 2 were purchased from Sigma. Mineral oil was obtained from Nacalai Tesque. The fluorescent phospholipid, Texas Red

1,2-dihexadecanoyl-sn-glycero-3-phosphoethanolamine, triethylammonium salt (Texas Red DHPE) was obtained from Molecular Probes. For the experiment on cell-free gene expression, a PROTEINscript-PRO Kit which included E coli S30 extract, T7 RNA polymerase, Master Mix, and amino acids excluding L-methionine was purchased from Ambion. L-Methionine was purchased from Nacalai Tesque and Plasmid pQBI T7 (5115 bp) encoding a T7 promoter and an rsGFP gene was obtained from Nippon Gene as template DNA. We used nuclease-free water from Otsuka Pharmaceutical Factory for the gene expression experiment.

We used a Nikon TE300 microscope and a Hamamatsu Photonics EBCCD camera and performed the measurement at ambient temperature (around 20 ℃). For the observation of GFP, we used an Olympus IX-70 microscope and an Andor iXon DV887 EMCCD camera equipped with a temperature controller (TOKAI HIT microscope incubation system). For laser tweezers, a Nd: yttrium-aluminum garnet (YAG) laser (SL902T, Spectron) with a $TEM_{00}$ beam at a wavelength of 1064 nm was used.

We prepared 1 mM DOPE in mineral oil by ultra-sonication for 90 min at 50 ℃ and used it within 24 hours. Next, 2 μL of aqueous solution was added to 200 μL of mineral oil containing DOPE in a test tube and mixed by a micropipette for a few seconds to form CWDs. To observe these droplets in the bulk oil phase, we used a chamber made of silicone rubber on a hydrophobic cover glass, as shown in Fig. 1A. Individual CWDs can be manipulated by laser tweezers, where a CWD feels repulsive force from the laser focus. Figure 1B exemplifies phase contrast and fluorescent microscopic images of a CWD formed in the presence of DOPE containing 1% Texas Red DHPE. It is clear that a phospholipid layer surrounds the interface of the droplet. Figure 2A shows the procedure used to fuse two CWDs, where the droplet on the left is pushed toward the one of the right by laser tweezers. In this experiment, the fusion started 4.2 seconds after the forced contact, and the entire fusion process was complete in less than 0.03 seconds, i.e., within the video frame-rate. The volume of the new fused CWD is essentially the same as the sum of those of the two original CWDs. Figure 2B shows the diffusion of the fluorescent dye Fura 2, originally encapsulated in the droplet on the left, into the newly generated vesicle after coalescence, together with a spatio-temporal representation. At the very instant of fusion, the bright fluorescent region is localized in the left hemisphere (0.00 s) and then diffuses into the remaining part (0.00 - 0.30 s).

Figure 2 C shows fluorescent microscopic images of diffusion on the membrane surface after the coalescence of CWDs, where the phospholipid layer of the droplet on the left contains 1% Texas Red DHPE. The wave front of the diffusion on the two-dimensional surface is visualized, and shows that the speed of diffusion is rather slow, with more than three minutes needed for complete mixing.

Next, we examined the applicability of a CDW as a microscopic biochemical reactor. We adopted a simple enzymatic reaction between the enzyme esterase and calcein-AM as a substrate, where the esterase dissolves the acetoxymethyl (AM) group from calcein-AM and the reaction product, calcein, emits fluorescence proportional to its concentration. Figure 3 shows fluorescence microscopic images of a time series of a coalesced droplet (upper) and a graph of the time-development of fluorescence intensity (lower) after the fusion of two CWDs containing 50 μM calcein-AM and 2.5 μg/ml esterase, respectively. The initial increase in the reaction rate corresponds to the rate of formation of the substrate and enzyme complex, since the components have been mixed for less than 1 s after fusion, as indicated in Fig. 2B, and does not reflect the apparent kinetics above the order of a second. In the intermediate time-region with a nearly constant reaction rate, the concentration of the enzyme-substrate complex is practically constant. The decline in the late-state kinetics is attributable to a deficiency of the reaction substrate. Thus, it is apparent that a CWD can act as a micro biochemical reactor, with no inhibitory effect on the enzyme activity.

We next measured the production of a protein, GFP, in a single CWD, using cell-free expression under temperature control at 37 ℃. Figure 4 shows the fluorescence microscopic images and time-development of the GFP expression reaction within the CWD. To perform the expression reaction as in Fig. 4, we prepared two kinds of CWDs that encapsulated different components; one contained E coli S 30 extract, Master Mix, 100 μM each amino acid, and 2.5 U/μl T7 RNA polymerase as developed by Zubay[20]; and the other contained 0.25 μg/μl pQBI T7 plasmid DNA. The fluorescence intensity, i.e., the amount of GFP produced, increased with time, and attained a maximum value at around 120 min. Finally, the fluorescence intensity decreased due to spontaneous degradation.

In the present study, we used DOPE as a phospholipid because it is known to be a major component in the inner cytoplasmic membrane.[21] However, it has been reported that DOPE usually forms a inverted hexagonal structure and does not form stable vesicles in aqueous solutions at usual physiological solution conditions.[22] To the best of our knowledge, no previous studies have successfully used DOPE to form a model membrane system Thus, our report describes a new useful methodology for preparing a model membrane system with DOPE.

We now discuss the mechanism of the fusion of CWDs from a physicochemical point of view. As shown in Fig. 2 A, the coalescence of two CWDs occurs easily through mutual contact with the use of laser tweezers. In contrast, it has been reported that it is rather difficult to produce the fusion of two cell-sized liposomes, as giant vesicles. To fuse giant vesicles, a chemical substance, such as a surfactant or ultivalent Lanthanum ion is added to the vesicle solution to destabilize the bilayer.[23,24] The other known method for vesicle fusion is the application of a rather strong electric field pulse.[13] With these methodologies, it is practically difficult to fuse a desired pair of vesicles. We would like to address the benefits of a CWD as a small biochemical reactor. Considering the small size of a CWD, we can expect rapid "mixing" without any stirring procedure. The mixing time of the substances encapsulated in the separate CWDs can be estimated by considering simple diffusion process

$$\Delta t_{diff} \sim r^2 / 4D \approx 1[s] \tag{1}$$

where $D$ is the diffusion constant ($D = k_B T / 6\pi\eta a$), $k_B$ is the Boltzmann constant, $T$ is temperature and $a$ is the radius of the molecule ($\sim 10^{-9}[m]$, for a typical molecule). It can be expected that convection would also contribute to mixing. In our actual experiments, the "mixing" time is less than 1 s, as indicated in Fig. 2 B.

Finally, we will discuss the diffusion of lipid molecules on a two-dimensional surface, as in Fig. 2 C. The diffusion constant on a two-dimensional surface is given as $k_B T / 4\pi\eta a$ ( $\sim 10^{-12} [m^2/s]$ in the case of reconstituted phospholipid membrane[25]). By considering a characteristic length of diffusion on the order of several micrometers, the time for the diffusion of phospholipids after coalescence is estimated to be $\Delta t_{diff}^{lipid} \sim 10^1 [s]$. This time-scale of diffusion has been confirmed in a past experiment on cell-sized bilayer phospholipid vesicles, where diffusion was shown to be complete within about 7 seconds.[12] On the other hand, this estimate is one order of magnitude less than our experimental data (180 seconds in Fig. 2 C). We measured the viscosity of mineral oil under the conditions in the present study, and found that its viscosity is 13 times grater than that of water.[26] Thus, we concluded that the slow diffusion rate in a CWD is due to the large viscous friction from the bulk oil phase.

The present results suggest that CWDs can serve as useful micro-reactors and also as simple models of living cells. The new experimental methodology reported here may contribute to a deeper understanding of the mechanism of biochemical reaction networks in microscopic environment in living cells.

This work was supported by a Grant-in-Aid for Scientific Research on Priority Areas (No.17076007) "System Cell Engineering by Multi-scale Manipulation" from the Ministry of Education, Culture, Sports, Science and Technology of Japan. M. Hase and A. Yamada were financially supported by Kyoto University Venture Business Laboratory and a Sasakawa Scientific Research Grant from The Japan Science Society, respectively. T. Hamada was supported by a Research Fellowship from the Japan Society for the Promotion of Science for Young Scientists (No. 16000653). We are grateful to Prof. A. Mizuno and Prof. S. Katsura of Toyohashi University of Technology for their kind advice. We also thank Mr. S. Watanabe for his help in the viscosity measurements.

**Figure Captions**

**Fig. 1 A:** Schematic illustration of the experimental system. The desired droplet, a CWD, is transported by use of laser tweezers, where the CWD experiences a repulsive force from the laser focus. **B**: Phase contrast (a) and fluorescence (b) microscopic images of a CWD in an oil layer containing DOPE with 1 % Texas Red DHPE. The images indicate that a phospholipid layer is formed around the CWD, most probably as a lipid monolayer.

**Fig. 2 A**: Phase contrast images of the fusion of two droplets by the use of laser tweezers. At 6.03 s, the frame shows the overlap of different images corresponding to those before and after fusion. This indicates that fusion is complete within the video frame rate, 0.03 s. **B**: Fluorescent microscopic images of the bulk diffusion of 50 μM Fura 2 encapsulated in the CWD on the right into that on the left which contains water before (a) and after (b) coalescence. (c) Spatio-temporal image of the centerline of the resulting CWD. **C**: Fluorescent microscopic images of the diffusion of fluorescent phospholipids on the CWD surface, from the CWD on the right into that on the left before (a) and after (b) coalescence. (c) Spatio-temporal image of the circular arc of the CWD.

**Fig. 3:** Fluorescence microscopic images of a single CWD undergoing the reaction of calcein-AM with esterase, as visualized by the emission of calcein as the product. ~~the~~ The graphs show the time-dependent change in fluorescence intensity (I), together with the velocity of the reaction represented as the time derivative of I. t = 0 is taken as the moment of the fusion of droplets containing the substrate and enzyme, respectively.

**Fig. 4** Fluorescence microscopic images of a CWD undergoing gene expression, i.e., the production of GFP, at 0 and 3 hours after the coalescence of droplets containing DNA and expression medium kept at 37 ℃. The graph shows the time-development of the fluorescence intensity of GFP.

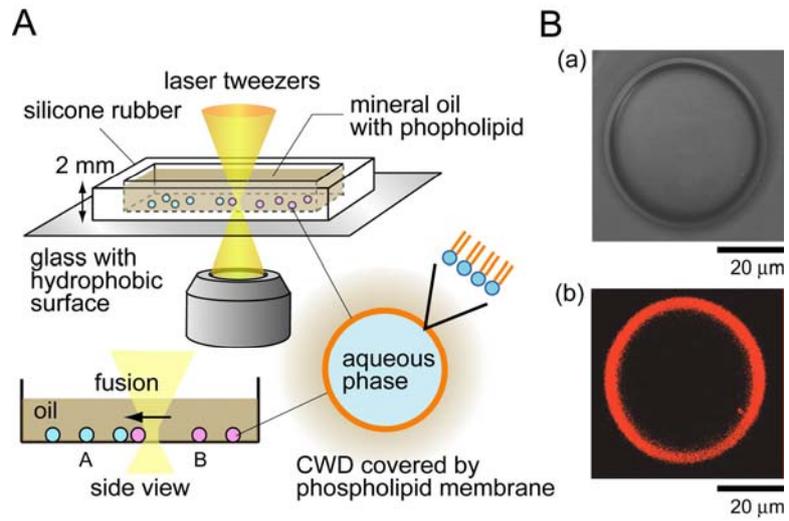

**Figure 1**

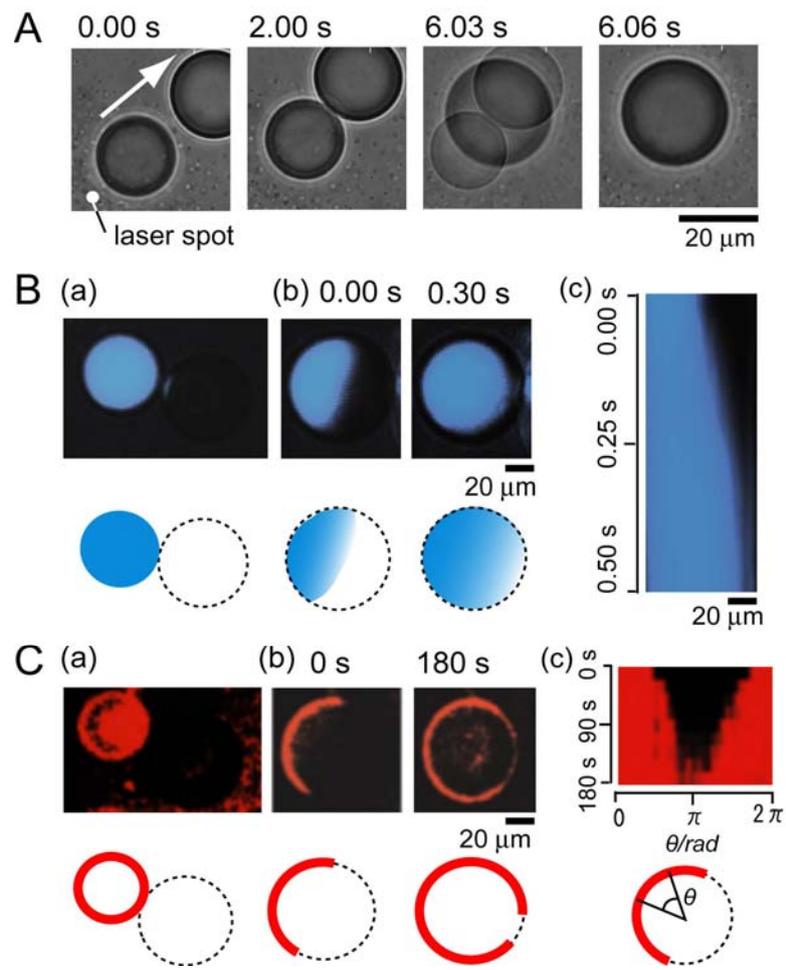

**Figure 2**

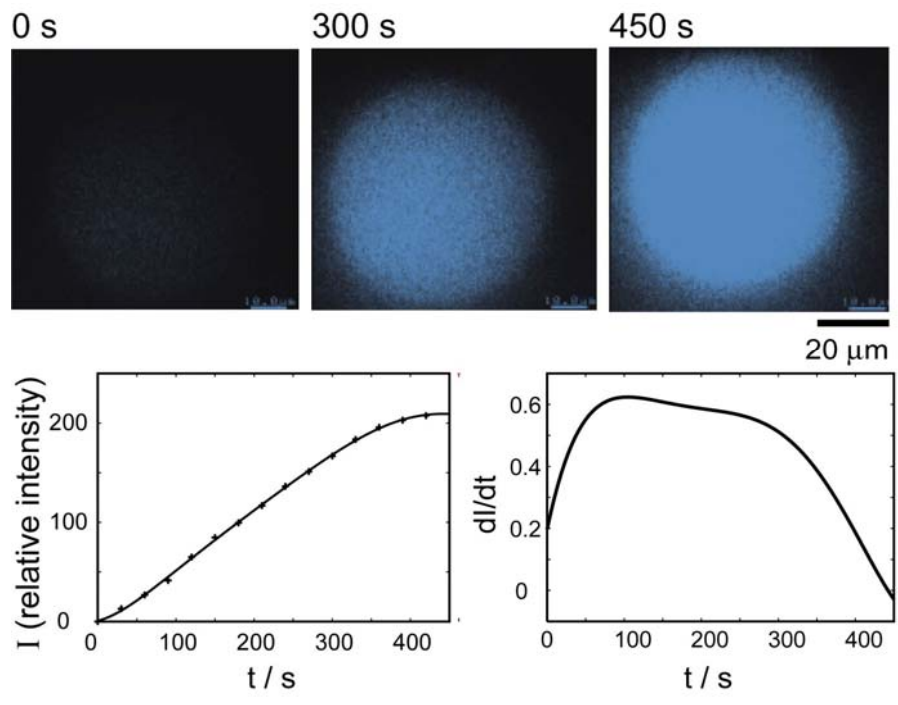

**Figure 3**

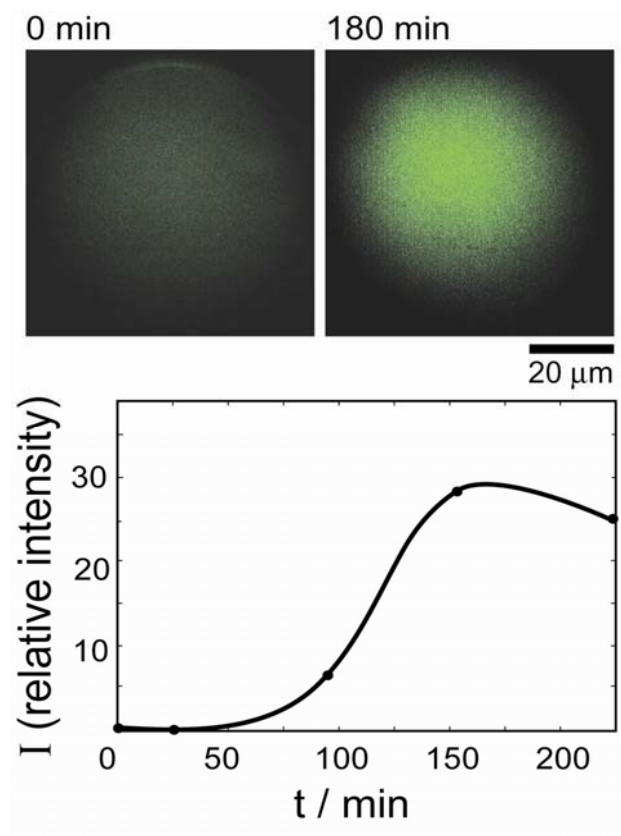

**Figure 4**